\documentclass[10pt,conference]{IEEEtran}

\usepackage{url}
\usepackage{hyperref}
\usepackage{multirow}
\usepackage{listings}
\usepackage{courier}
\usepackage{color}
\usepackage{multicol}
\usepackage{etoolbox}
\usepackage[font=small,labelfont=bf]{caption}
\usepackage[utf8]{inputenc}
\usepackage[T1]{fontenc}
\usepackage[english]{babel}
\usepackage[cmex10]{amsmath}
\usepackage{fancyvrb}
\usepackage{xcolor}

   \usepackage[pdftex]{graphicx}     
   
   \graphicspath{{../pdf/}{../jpeg/}{./image/}}

   \DeclareGraphicsExtensions{.pdf,.jpeg,.png,.jpg}

\begin{document}

\title{\huge A Framework for Lattice QCD Calculations on GPUs}

\author{
    \IEEEauthorblockN{F. T. Winter\IEEEauthorrefmark{1}, M. A.
      Clark\IEEEauthorrefmark{2}, R. G. Edwards\IEEEauthorrefmark{1},
      B. Jo\'o\IEEEauthorrefmark{1}}
    \IEEEauthorblockA{\IEEEauthorrefmark{1}Thomas Jefferson National
      Accelerator Facility, Newport News, VA, USA}
    \{fwinter,edwards,bjoo\}@jlab.org
    \\
    \IEEEauthorblockA{\IEEEauthorrefmark{2}NVIDIA Corporation, 2701 San Tomas Expressway, Santa Clara, CA 95050, USA
    \\mclark@nvidia.com}
}

\maketitle
\thispagestyle{plain}
\pagestyle{plain}

\begin{abstract}

Computing platforms equipped with accelerators like GPUs have proven to provide great computational power. However, exploiting such platforms for existing scientific applications is not a trivial task. Current GPU programming frameworks such as CUDA C/C++ require low-level programming from the developer in order to achieve high performance code. As a result porting of applications to GPUs is typically limited to time-dominant algorithms and routines, leaving the remainder not accelerated which can open a serious Amdahl's law issue. 

The lattice QCD application Chroma allows to explore a different porting strategy. The layered structure of the software architecture logically separates the data-parallel from the application layer. The QCD Data-Parallel software layer provides data types and expressions with stencil-like operations suitable for lattice field theory and Chroma implements algorithms in terms of this high-level interface. Thus by porting the low-level layer one can effectively move the whole application in one swing to a different platform. 

The QDP-JIT/PTX library, the reimplementation of the low-level layer, provides a framework for lattice QCD calculations for the CUDA architecture. The complete software interface is supported and thus applications can be run unaltered on GPU-based parallel computers. This reimplementation was possible due to the availability of a JIT compiler (part of the NVIDIA Linux kernel driver) which translates an assembly-like language (PTX) to GPU code. The expression template technique is used to build PTX code generators and a software cache manages the GPU memory. 

This reimplementation allows us to deploy an efficient implementation of the full gauge-generation program with dynamical fermions on large-scale GPU-based machines such as Titan and Blue Waters which accelerates the algorithm by more than an order of magnitude.

\end{abstract}

\section{Introduction}

Recent trends in HPC computer architecture are making massively
multicore parallelism and heterogeneity omnipresent.
Computer centers around the globe are deploying supercomputers utilizing
hybrid architectures accelerated with Graphic Processing Units (GPUs).

NVIDIA established CUDA as a parallel computing platform and
programming model which enables developers to fully exploit the
computational power of GPUs.
It supports heterogeneous computation where applications off-load 
parallel portions to the GPU.
Typically, CUDA-enabled computation is applied incrementally to
existing applications.

Mature scientific applications usually consist of huge source
codes and porting such applications to architectures like CUDA
includes code refactoring into off-load regions (kernels), dealing with
explicit memory management and data-layout optimizations.
This makes applications porting time-consuming, tedious and
error-prone.
Therefore application porting efforts are often limited in scope to
central routines.

Lattice Quantum Chromodynamics (LQCD) is a good example where this
porting practice is often applied.
LQCD calculations are usually divided into two main parts:
The Hybrid Monte Carlo (HMC) gauge field generation part typically
requiring capability computing and the analysis part in which the
physical observables are determined. 
For the latter capacity computing will typically suffice.
The Chroma application suite \cite{Edwards:2004sx} for LQCD
calculations includes both, the post-Monte Carlo analysis and gauge
field generation part.

A common operation to both parts is finding the solution to a large linear
system of equations, where the matrix in question is the sparse system
that arises from discretizing the Dirac operator.
Depending on the simulation parameters, e.g. the quark masses, the
linear solves are the operations where most of the compute time is
spent and where porting efforts typically focus on.
An example of this strategy is the QUDA library \cite{Clark:2009wm},
which provides CUDA-accelerated linear solvers for common Dirac
operator discretizations but whose support beyond this is presently
limited.

QUDA has been successfully applied to the post-Monte Carlo analysis
phase which is often dominated by the linear solves.
However, the gauge field generation part is more diversified and just
adding drop-in libraries like QUDA opens a serious Amdahl's law
issue.
E.g., when deploying the Chroma RHMC ($40^3\times 256$, 2+1 flavors 
of dynamical quarks, $m_\pi \approx 230$ MeV) executing on the CPU
with acceleration only provided through QUDA we measure a speedup
factor between
$\sim2.2$ (128 GPUs/CPU sockets) and
$\sim1.8$ (800 GPUs/CPU sockets)
for our production HMC simulations on Blue Waters.
The relatively large fraction of the computation which is not part of
a linear solve plus the incurred data transfer between GPU and CPU
memories render the gauge field generation part of LQCD calculations
not amenable for accelerating solely via the linear solves.

Chroma, however, allows to pursue a different porting strategy.
This application is part of a layered software architecture which
decomposes its functionality into logical groupings of software
components where separate layers or components communicate with each
other through clearly defined interfaces.
The QCD Data Parallel (QDP++) software layer provides data types
and operations suitable for lattice field theory.
This is the low-level layer which addresses all characteristics of a
particular hardware architecture and hides them from the high-level
layer.
The application layer implements algorithms in terms of the
domain-specific data types and operations provided by this interface.
Thus, Chroma implements LQCD domain science in a target machine
independent way and is as such easier to maintain and extend.
It represents a major investment made in software development and its
source code consists of over 400,000 C++ code lines.

As a result of the layered structure of the software architecture
Chroma can be ported to the CUDA architecture by providing a
reimplementation of the low-level layer.
This endeavor would have an immediate, large impact as a whole LQCD
application suite would instantly tap into the computational power of
acceleration as opposed to the traditional incremental application
porting.
With all computation executing on the accelerator this also would
effectively alleviate the effects of Amdahl's law when interfacing to
external libraries such as QUDA.

Deploying Chroma with our reimplementation of the low-level layer for
CUDA we measured
a speedup factor of 
$\sim11.0$ ($128$ GPUs/CPU sockets) and 
$\sim3.7$ ($800$ GPUs/CPU sockets) 
for our production RHMC simulations on Blue Waters.

It is worth noting that the design of the data-parallel level as found
in our software architecture is by no means restricted to lattice
field theory.
This level implements an (embedded) data-parallel language similar to
High Performance Fortran (HPF) \cite{Loveman:1993:HPF:613767.613797}.
We identified the following design elements to be of enabling
character for our work:
The definition of a data-parallel container, 
whole-array elemental operations and 
support for expressions.
These are only a few characteristics of data-parallel languages which
might be found in other application domains.

In the following we refer to the QDP-JIT/PTX library as the ``new
implementation'' and to the QDP++ library as the ``original
implementation''.

While the expressions which are supported by the interface have a
positive impact on science productivity they represent a significant
challenge to the library developer when implementing the interface for
the CUDA architecture.
To date no GPU programming framework is capable to automatically
generate accelerated code from expression templates employed in the
host program.
In order to accelerate expressions advanced annotation or code
refactoring, i.e. code separation into CUDA kernel routines, is
needed.

However, the structural information about the expressions is encoded
in the templates and is accessible at compile time and can be turned
into code generators \cite{DBLP:journals/corr/cs-PL-9810010}.
Thus, we employed a modified version of expression templates in our
new implementation so that kernel code generators are built.
The code generators, when triggered at runtime, generate CUDA kernels
in the PTX language \cite{PTX31}.
Those are subsequently translated by the NVIDIA JIT compiler to
executable GPU code.
This procedure was found to execute smoothly.
Neither the code generation, nor the subsequent compilation by the
compute compiler generate a significant amount of overhead as we shall
demonstrate later on.

Due to CUDA's off-loading execution model user data can be accessed by
host and kernel code.
Thus the developer is required to explicitly issue memory copy
instructions according to the data access pattern of the program.
Given the large source code of Chroma, adding memory transfer
instructions explicitly would be laborious and error prone to the
point of impracticality.
To meet the requirements of this dynamic access pattern our new
implementation adds comprehensive memory management to the library.
Again, we use the expression templates, more precisely the runtime
state information stored therein, namely references to data fields.
This allows us to make all referenced data fields available in the GPU
memory domain prior to launching a kernel.
Due to the dynamic nature of these memory accesses we implemented a
``cache'' for GPU memory which can spill least recently used data
fields to CPU memory in order to allocate new fields on the GPU.
As a result CUDA memory management is completely automated and
altering or annotating the application source code is not needed.

An optimization technique which we applied was changing the data
layout for data fields to suit the GPU memory access patterns.
We employed a structure of arrays (SoA) data layout such that memory
accesses are coalesced.
We added a low-overhead auto-tuning step for each compute kernel.
This auto-tuning determines experimentally a thread block size,
maximizing kernel performance.
Another optimization that we applied is overlapping of communication
and computation on stencil-like operations.

We found that on the NVIDIA Kepler architecture the generated compute
kernels sustain 79\% of the peak memory bandwidth, which is the
appropriate metric of efficiency, since the kernels in LQCD
computations are memory bandwidth bound.

Our new implementation has allowed us to deploy the full Chroma
gauge-generation program in production on large scale GPU-based
machines such as Titan at the Oak Ridge Leadership Computing Facility (OLCF),
and Blue Waters at the NCSA Petascale Computing Facility.
Chroma built on top of the QDP-JIT/PTX library combined with the
linear solvers from the QUDA library represent a high-performance
configuration for running HMC calculations on GPU-based systems.

\begin{table*}[t]
\centering
\caption{
  Data types in QDP++, \texttt{REAL} $\in$ \{\texttt{float}, \texttt{double}\}.
  The lower part of the table is not part of QDP++'s type system.
  These data types were added to support the Clover term in the Chroma application.
  \label{tab:types}
}
\begin{tabular}{ |p{0.15\columnwidth}|p{1.4\columnwidth}|p{0.4\columnwidth}| }
\hline
symbol & definition & type alias\\
\hline
$\psi$ &   
\texttt{Lattice< Vector< Vector<   Complex< REAL >, 3>, 4> >} &
\texttt{LatticeFermion} \\
$U$ &      
\texttt{Lattice< Scalar< Matrix<   Complex< REAL >, 3>   > >} &
\texttt{LatticeColorMatrix} \\
$\Gamma$ & 
\texttt{Lattice< Matrix< Scalar<   Complex< REAL >   >, 4> >} &
\texttt{LatticeSpinMatrix} \\
\hline
$A_\texttt{diag}$ & 
\texttt{Lattice< Component< Diagonal<   Scalar< REAL > > > >} & 
\textsl{n.a.}\\
$A_\texttt{tria}$ & 
\texttt{Lattice< Component< Triangular< Complex< REAL > > > >} & 
\textsl{n.a.}\\
\hline
\end{tabular}
\end{table*}

This paper is organized as follows.
We summarize the necessary background in Sec.~\ref{sec:bg}, 
giving an overview of the scientific domain, 
discussing the QDP++ application framework, stencil operations and 
the CUDA architecture.
We describe our code generation process in Sec.~\ref{sec:codegen},
giving details on JIT compilation, building code generators and our
PTX code generator.
Sec.~\ref{sec:CCM} and Sec.~\ref{sec:overlapping} discuss the automatic
memory management and the overlapping of communication and kernel
computation respectively.
We describe the functions discussed in the experimental results and
kernel auto-tuning in Sec.~\ref{sec:functions} and
Sec.~\ref{sec:tuning}.
We show our numerical results in Sec.~\ref{sec:results} and
Sec.~\ref{sec:related} details on related work.
We summarize and conclude in Sec.~\ref{sec:conclusions}.

\begin{table}
\centering
\caption{Test functions for our benchmarking runs.
Notation for gauge, fermion fields as in Table \ref{tab:types}.
\label{tab:funcs}}
\begin{tabular}{ |p{0.15\columnwidth}|p{0.5\columnwidth}|p{0.2\columnwidth}|}
\hline
Test & Expression & flop/byte (DP) \\
\hline
\texttt{lcm}      & $U_1=U_2*U_3$ & 0.458 \\
\texttt{upsi}     & $\psi_1=U_1*\psi_2$ & 0.5 \\
\texttt{spmat}    & $\Gamma_1=\Gamma_2*\Gamma_3$ & 0.62\\
\texttt{matvec}   & $\psi_0=U_1*\psi_1+U_1*\psi_2$ & 0.64\\
\hline
\texttt{clover}   & $\psi_0=A*\psi_1$ & 0.525 \\
\hline
\end{tabular}
\end{table}

\section{Background}

\label{sec:bg}

\subsection{Lattice QCD}

\label{sec:latticeqcd}

Lattice QCD (LQCD) is the lattice-regularized formulation of Quantum
Chromodynamics (QCD), the theory of the strong interaction between
quarks and gluons. 
In LQCD, the continuum theory is typically discretized on a Euclidean
space-time embedded in a hypercubic, 4-dimensional lattice and the QCD
path integral over the quark and gluon fields is performed using Monte
Carlo methods.
LQCD is the most successful, systematic approach to calculate
properties of the non-perturbative regime of QCD and is an important
tool for nuclear and high-energy physics.
A full description of LQCD is beyond the scope of this paper and we
refer the reader to the many excellent references available such as
\cite{Creutz:lattice}.

In terms of implementation on a computer we note that there are
two primary data ``types'' in QCD calculations namely the gluon
(Gauge) fields which are typically denoted $U_\mu(x)$.
These variables are complex SU(3) matrices typically ascribed to the
links between lattice sites.
Here the coordinate $x$ refers to the lattice site from which the link
emanates and $\mu$ is a forward direction $\mu \in [0..N_d-1]$ with
$N_d$ being the number of space-time dimensions.
Quark fields are discretized on lattice sites and can be denoted as
$\psi_{C,S}(x)$ where $C$ and $S$ are color and spin indices
respectively.
Hence, a framework, such as QDP++, needs to provide data types which
have indices in the space-time and the internal symmetry (spin, color)
domains.
Finally we note that nearly all lattice types are represented with
complex numbers.

\subsection{The QDP++ Application Framework}

\label{sec:qdp}

In the following we will introduce concepts of QDP++ which will be
necessary in the remainder of the paper.
First we will give a very brief overview of the framework and
introduce central data types and operations defined in the interface.
Finally we will detail some aspects of its implementation.

A central data type is the data-parallel container \texttt{Lattice}
which implements the space-time structure of Lattice QCD.
This container ascribes its elements logically to grid points on an
$N_d$-dimensional hypercubic lattice.
This container is said to live on the ``outer level'' of the data type
hierarchy and node parallelization, typically with MPI, is implemented
on this level.
Thus, each node (or rank) maintains a ``sub-grid'' of the
global lattice.

The internal symmetry (spin, color) domains live on the ``inner
levels'' of the hierarchy.
The composition of data type levels is implemented with template
nesting. Thus a complete data type is composed of four levels, named
after the QCD index spaces
$$
\texttt{Lattice} \otimes
\texttt{Spin} \otimes
\texttt{Color} \otimes
\texttt{Complex}.
$$
The interface defines class templates which can be used on inner index
spaces as building blocks of the data type system, e.g.
\texttt{Scalar}, \texttt{Vector}, \texttt{Matrix},
\texttt{Complex} and \texttt{Real}.

For example, a lattice fermion $\psi$ (we omit the indices here) is a
``spin-color vector'', i.e. it has vector structure in spin and color
space, and is defined as shown in Table~\ref{tab:types} (upper part).
The table shows other commonly used data types.
For the user's convenience type aliases are defined for frequently
used data types, such as \texttt{LatticeFermion}.

The QDP++ interface defines its operations in operator infix form,
i.e. C++ expressions, which can be applied to data-parallel
\texttt{Lattice} instances.
For example, a multiplication of
\texttt{Scalar}$\times$\texttt{Vector} in spin space,
\texttt{Matrix}$\times$\texttt{Vector} in color space and
\texttt{Complex}$\times$\texttt{Complex} in reality space can be
written as:

\begin{verbatim}
psi = u * phi;
\end{verbatim}
with 
\texttt{u} of type \texttt{LatticeColorMatrix} and
\texttt{psi,phi} of type \texttt{LatticeFermion}.
N.B. the operations are implicitly data-parallel, i.e. no loop over the
lattice site index has to be written explicitly.
The library implements the infix form with expression templates,
more precisely with the PETE library \cite{exprtemplates, pete}.

\subsection{Shift Operations}

In Lattice QCD one is often interested in writing discretized
derivative operators such as the Laplacian or Dirac operator.
In order to build such operators in a data-parallel language one
typically needs stencil-like operations. 
The QDP++ interface defines so-called \texttt{shift} operations as
building blocks for stencil-like computations.
These are data-parallel operations that displace the underlying grid
points in the specified dimension and direction by one grid point,
similar to the circular shift operations in HPF 
\cite{Loveman:1993:HPF:613767.613797}.

For example, Figure~\ref{fig:1dim} shows the gauge covariant form of a
simple 1-dimensional nearest-neighbor discretization of a partial
derivative in $\mu$ direction.
\begin{figure}[t]
\small
\begin{Verbatim}[frame=single]
multi1d<LatticeColorMatrix> u(Nd);
LatticeFermion psi,phi;

psi = u[mu] * shift(phi , mu , FORWARD) + 
   shift( adj( u[mu] ) * phi , mu , BACKWARD);
\end{Verbatim}
\normalsize
\caption{Simple nearest-neighbor discretization of a partial
  derivative using the QDP++ interface. \label{fig:1dim}}
\end{figure}
The vector-like container \texttt{multi1d} is provided for convenience
and (in this example) bundles together the gauge links in different
dimensions.
Shifting in backward direction $-\mu$ requires multiplication with the
Hermitian adjoint of the SU(3) matrix.

Operations including stencil operations provide the opportunity to
overlap communication and computation.
Communication refers here to inter-MPI rank data transfers.
This optimization is implemented in the QDP-JIT/PTX library as
described in Sec.~\ref{sec:overlapping}.

\subsection{CUDA Architecture}

\label{sec:ptx}

The CUDA architecture is a parallel computing platform and programming
model created by NVIDIA for use with their GPUs.
CUDA C/C++ provides a small set of extensions to standard programming
languages, like C/C++, that enables developers to implement parallel
algorithms for GPUs.

The Parallel Thread Execution (PTX) \cite{PTX31} is an intermediate
assembler language between high-level CUDA C/C++ code and GPU machine
code.
It is a machine-independent virtual machine and instruction set
architecture which was made publicly available by NVIDIA.
The low-level language interface to kernel programming is a key
enabling technology for this work as it has allowed us to implement
compute kernels directly in assembly language and interface to a
JIT compiler.

Figure~\ref{fig:cuda} shows a schematic of the compilation steps and
program representations involved for the GPU kernel part in a typical
CUDA C/C++ development chain.
Kernel code is extracted from the source code by the NVCC compiler and
translated to the PTX language.
The NVIDIA compute compile driver (part of the Linux kernel driver)
optimizes and translates the PTX kernel program to GPU machine code.

\section{Code Generation}

\label{sec:codegen}

The following section describes how the QDP-JIT/PTX library generates
GPU kernel code from data-parallel expressions.
GPU kernel code is generated using the PTX programming language which
we will also refer to as the ``secondary language'' (translated by the
JIT compiler) as opposed to the application or host programming
language (C++).

\subsection{JIT Variables}

\begin{figure}[t]
\framebox[\columnwidth]{
\begin{minipage}[c][1cm][c]{\columnwidth}
\begin{center}
CUDA C/C++
$\xrightarrow{\mbox{\small{NVCC}}}$
PTX
$\xrightarrow{\mbox{\small{Linux driver}}}$
GPU code
\end{center}
\end{minipage}
}
\vspace{-0.05cm}
\caption{\label{fig:cuda}Program representations and compilation steps
 in the NVIDIA CUDA C/C++ development chain. Our compute kernels are
 implemented directly with the PTX programming language.}
\end{figure}

First we review the concept of JIT variables.
These are quite similar to the familiar variables used in computer
programs as they have a type and symbolic names in the secondary
language. 
An important difference, however, is that they are not first-class
objects of the host language; they cannot be stored in an usual
variable, neither can they be instantiated at runtime in the
execution context of the host program.
JIT variables abstract storage locations of the JIT execution context
and they are made available to computational manipulation in the host
language through reification as manipulable variables in the host 
language which are often called ``JIT values''.
Manipulating JIT values allows to build computer programs in a
secondary language interfacing to a JIT library which provides the
arithmetic operations.

\subsection{JIT Data Views}

We introduce ``JIT data views'' (or simply JIT views) in order to
establish a connection between the host program's and the JIT
program's execution context.
We need to make the abstract C++ container types,
i.e. \texttt{Lattice}, visible to the JIT context.
In other words, we need to issue the appropriate load and store
instructions from and to memory in order to make data values available
for arithmetic manipulation in the register file.

Given a sequence container of (structured) elements which are
contiguously stored in memory, JIT views provide the means to
calculate the index position of a particular data word inside the
container.
Thus, together with the base memory address of the container's raw
data array the index position can be used to calculate the memory
offset for the load and store instructions.

The index position is calculated based on a data layout function.
This function can be any arithmetic function of the index domains and
indices of the domain-specific data types.
Since the target architecture is CUDA we use a data layout function
that results in coalesced memory accesses:
$$
  I(i_V ,i_S,i_C,i_R) = ((i_R * I_C + i_C)*I_S+i_S)*I_V+i_V
$$
where $I_V$,$I_S$,$I_C$,$I_R$ are the index domains for, and
$i_V$,$i_S$,$i_C$,$i_R$ are the indices within these domains for
the space-time, spin, color and complex component respectively.
The coalescing condition can be easily read off from the above
function when assuming thread-parallelisation takes place on the
space-time index domain, i.e. adjacent threads (thread number $i_V$)
access adjacent memory locations.

\subsection{The Code Generation Process}

In the following we describe the building process of code
generators as implemented in the QDP-JIT/PTX library.
The general concept of generating code via C++ expression templates
was pointed out in \cite{DBLP:journals/corr/cs-PL-9810010}.
Here we will review the basic ideas behind this mechanism as used in
our original implementation, and then outline the changes we've
employed in our new implementation.

\begin{figure}[t]
\centering  
$U_\mu(x)\psi(x+\hat\mu) \, + \, U_\mu^\dagger(x-\hat\mu)\psi(x-\hat\mu)$\\
\vspace{3mm}
\includegraphics[width=1.0\columnwidth]{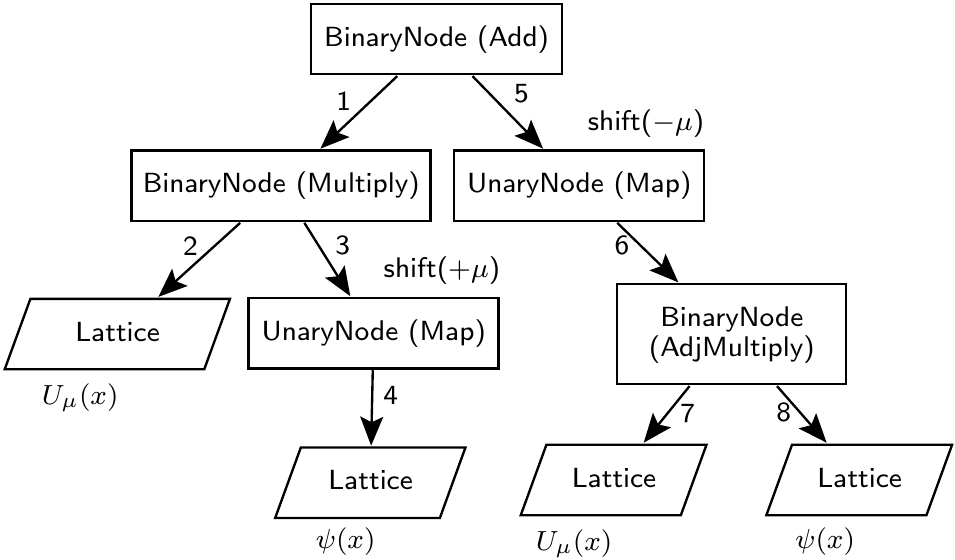}
\vspace{-5mm}
\caption{\label{fig:tree}
Example of an AST generated by PETE. 
Inner tree nodes represent operations.
Leaf tree nodes refer to data fields. 
The numbers give the visiting order when ``walking the tree''.}
\end{figure}

Expression templates are usually employed in vector libraries with
infix operator form in order to efficiently evaluate C++ expressions.
Overloaded operators return so-called ``proxy objects'' which
represent parts of the original expression, i.e. unary/binary
operations (unary minus, multiplication, etc.).
During template instantiation the compiler uses template nesting
as the composition operation for proxy objects. 
The nesting property of proxy objects gives the expression templates a
tree structure and the final class template representing the
expression can be visualized graphically as a tree, similar to an
Abstract Syntax Tree (AST).

Figure~\ref{fig:tree} shows an example of an AST for the derivative
operator shown in Figure~\ref{fig:1dim}.
Operations (\texttt{BinaryNode}) are represented as inner tree nodes
($+$,$*$,\texttt{Map},\texttt{AdjMultiply}) whereas data fields
(\texttt{Lattice}) are represented as tree leaf nodes.
The unary map operator implements the shift operations.

In order to evaluate the expression for a particular array index
given as a variable $i_V$, traditional expression templates traverse the
AST in depth-first order, following the numbers as shown in
Figure~\ref{fig:tree}, yielding runtime function calls to the operations
at the inner nodes (typically inlined by the compiler).
Visiting the tree leaf nodes returns the $i_V$-th vector element of the
container.
The loop over the vector index domain is ``pulled outside'' of the
expression, effectively eliminating the vector-sized temporaries
typically incurred in naive C++ operator overloading.
This unparsing process is also called \textsl{walking the AST}.

In analogy to traditional expression templates the QDP-JIT AST
unparser walks the AST, however, instead of yielding runtime function
calls to the operations, the unparsing process interfaces to a code
generator and yields code that, when executed, generates code in the
PTX language for that particular operation.
For example, the simple expression $a=b+c$ where $a$, $b$ and $c$ are 
vector containers the following (pseudo-) code would be generated

\footnotesize
\begin{verbatim}
jit_assign( View(a).elem(i_V),
    jit_add( View(b).elem(i_V), View(c).elem(i_V) ))
\end{verbatim}
\normalsize

Here, \texttt{jit\_assign} and \texttt{jit\_add} are calls to a
code generator (generating PTX in our case) and
\texttt{View::elem(i\_V)}
creates the earlier introduced JIT views into the vector container
pointing at the $i_V$-th element.
For the CUDA architecture, the loop over the site index is implemented
by CUDA thread parallelisation by setting $i_V$ to the thread index.

\subsection{PTX Code Generator}

\label{sec:ptxgen}

The last missing piece in the QDP-JIT/PTX code generation chain is the
PTX code generator which we will describe below.

Our code generator supports the common arithmetic, bit manipulation
and comparison operations on floating-point and integer data types.
While C/C++ supports implicit type conversions, e.g. using mixed
precision within a single arithmetic statement is allowed, the PTX
language is more restrictive.
Legal PTX programs require the proper use of data type conversion
instructions.
In order to support PTX code generation from mixed precision
expressions we added implicit type promotion.
Thus data type conversion instructions are silently issued as needed.

Employing the PTX language for writing programs in scientific computing
has one major downside:
The mathematical functions, i.e. those from the C math library, are
not available.
Only a small subset of these functions, the so-called ``fastmath''
functions, are supported.
These are functions for which fast, but reduced accuracy
approximations are implemented directly on the GPU hardware.

As a work-around, we generated PTX kernel programs for the relevant
mathematical functions with CUDA's static compilation tools (NVCC).
Based on this set of kernels we manually created PTX subroutines for
each of the functions.
The code generator will silently issue calls to the appropriate
subroutine every time a mathematical function is requested.
The downside of providing pre-generated PTX implementations in such a
way is that their implementation is tied to a particular version of
the PTX language and as a result a recompilation is needed every time
the PTX standard changes.

An advantage of implementing GPU kernels in the PTX language is that
the NVIDIA JIT compiler translates the kernels very quickly.
JIT compilation overhead for our compute kernels was found to be
between 0.05 and 0.22 seconds on the 12k compute node as specified in
the experimental result section, Sec.~\ref{sec:results}.

\section{Automated Memory\\ Management}

\label{sec:CCM}

The CUDA architecture implements an execution model where the GPU
portion of the program code is off-loaded as kernels to the
accelerator while non-kernel code executes on the CPU.
Thus our data structures such as \texttt{Lattice} are accessible by
kernel and host code and it's part of the library developers'
responsibilities to issue the appropriate  memory transfer and
synchronization instructions well before the actual access.

The QDP-JIT/PTX library automates this process by using a software
implementation of a cache.
Prior to a kernel launch for a given expression, we walk its AST,
e.g. the one shown in Figure.~\ref{fig:tree}, extract the references
to data fields stored at the leaf tree nodes and ``cache'' (make
available in GPU memory) the according data fields.
All data fields referenced by that particular kernel are then
available in GPU memory.

Data fields are ``paged-out'' (copied to CPU memory) either when they
are accessed by CPU code or upon a caching event that cannot be
serviced immediately due to too many cached data fields.
The problem of when to page-out a particular data field is decided by
a ``least recently used'' spilling algorithm based on the timestamp
of the last reference from a compute kernel.

\section{Overlapping MPI Communication and Computation}

\label{sec:overlapping}

On distributed memory systems the shift operations introduce data
dependencies on off-node grid points.
This provides the opportunity to overlap off-node communication with
kernel computation.
Since the node-local sub-grid is logically shaped as a
$N_d$-dimensional hypercubic grid the face sites, i.e. the sites to
send off-node, can be easily determined for a shift operation in a
particular dimension and direction.

For a given expression which includes shift operations the local
sub-grid is partitioned into ``inner sites'' and ``face sites''.
Compute kernels gather data into a contiguous region of GPU memory
from where it's sent directly (MPI) to the destination node.
For MPI implementations that are not ``CUDA-aware'' the data is first
copied to CPU memory.
While off-node data is in transition the compute kernel is launched
on the inner sites and after data has been received the remaining
sites are evaluated.
We do not detail much further on the implementation as this technique
is well established, for example see
\cite{Babich:2011:SLQ:2063384.2063478,Babich:2010mu}.

Our implementation is limited to simple shift operations as used,
e.g., in first-order finite difference operators.
Operations involving next-to-nearest neighboring site communications,
i.e. ``shifts of shifts'', execute the inner-most shift operations in
a non-overlapping fashion.

\section{Test Functions}

\label{sec:functions}

Table~\ref{tab:funcs} lists the test functions that we selected for our
benchmark tests.
These functions are relevant to LQCD calculations as they are used in,
e.g., calculations with SU(3) gauge fields, fermionic force terms and
spin projections.
The functions were chosen to exercise different combinations of index
spaces.

\subsection{Custom User-Defined Functions}

In LQCD, the so-called clover term \cite{Sheikholeslami1985} can be
added to the Wilson Fermion action in order to achieve $O(a^2)$
discretization errors.
The clover term is a local operator defined
as
\begin{equation}
A(x) = 1/4 \, c_{\rm SW} \, \sigma_{\mu\nu} F_{\mu\nu}(x)
\end{equation}
where the field strength tensor $F_{\mu\nu}(x)$ acts in color space and
$\sigma_{\mu\nu}$ in spin space.

In our particular spin basis (of $\gamma_\mu$ matrices) the clover
term is Hermitian, and splits into a block diagonal form.
There are two blocks, corresponding to the first two  and the second
two spin components respectively of the overall four components
available.
For each spin components we also have 3 color components giving each
block dimension 6 overall.
We store each block as the 6 real numbers along the diagonal, and the
15 complex numbers making up the lower triangular sub-diagonal part of
the block. 
The upper (super-diagonal) part can be found by Hermitian conjugation
(transpose and complex conjugation) as needed.

However, this implementation of the clover term mixes spin and
color index spaces in a non-trivial way and hence typical
implementations address this term outside of the QDP++ library since
there calculations in different index spaces are strictly separated
from each other.
Our code generation process, however, supports user-defined operations
even if they mix the spin-color structure.
We added support for the clover term on the application level,
i.e. in Chroma.
In terms of the nested QDP++ template structure, we use the template
level reserved for the color indices to store the triangular part of a
single block, and employ the template level usually used for spin
indices for the index of the two separate blocks.
See Table~\ref{tab:types} (lower part) for how our implementation makes
use of the spin and color spaces for the clover term.

\section{Kernel Auto-Tuning}
\label{sec:tuning}

In general, the performance of GPU kernels is impacted by factors such
as how the computation is organized into threads, and groups of
threads; in particular on the number of threads in a group, known as
the ``thread block size''.
However, the compute kernels generated by our code generator are
simple streaming kernels, which have no thread block communication and
their performance should have only a weak dependence on the thread
block size as long as enough threads are in flight.
Empirically we found, as expected, that on the Kepler architecture
running with a thread block size of at least 128 our streaming kernels
achieve the highest performance.

Running all the generated kernels with the same fixed thread block
size will likely not result in the most optimal performance, since
different kernels will have different resource requirements (e.g. in
terms of registers). Thus the same block size may result, for
example, in different levels of GPU utilization for different
kernels. Indeed some kernels may even exhaust resources and fail to
launch altogether.
Therefore we tune the thread block size automatically on a per kernel
basis.

Our auto-tuning strategy is implemented as follows: 
First we try to launch a given kernel with the maximum thread block
size allowed for the GPU in question (we use 1-dimensional blocks,
thus $2^{10}$ for Kepler) and, if that fails, re-try, having reduced
the thread block size by a factor of 2 until the launch succeeds.
Once successfully launched, consecutive launches ``probe'' smaller
block sizes until the execution time increases significantly
(arbitrarily we use 33\%).
The ``best configuration'' would then be used for all consecutive
launches.
We found that this auto-tuning method comes with very little
overhead:
No kernels are launched solely for the purpose of tuning; kernel
tuning is carried out on the payload compute launches.

\begin{figure}[tb]
\centering
\includegraphics[width=1.0\columnwidth]{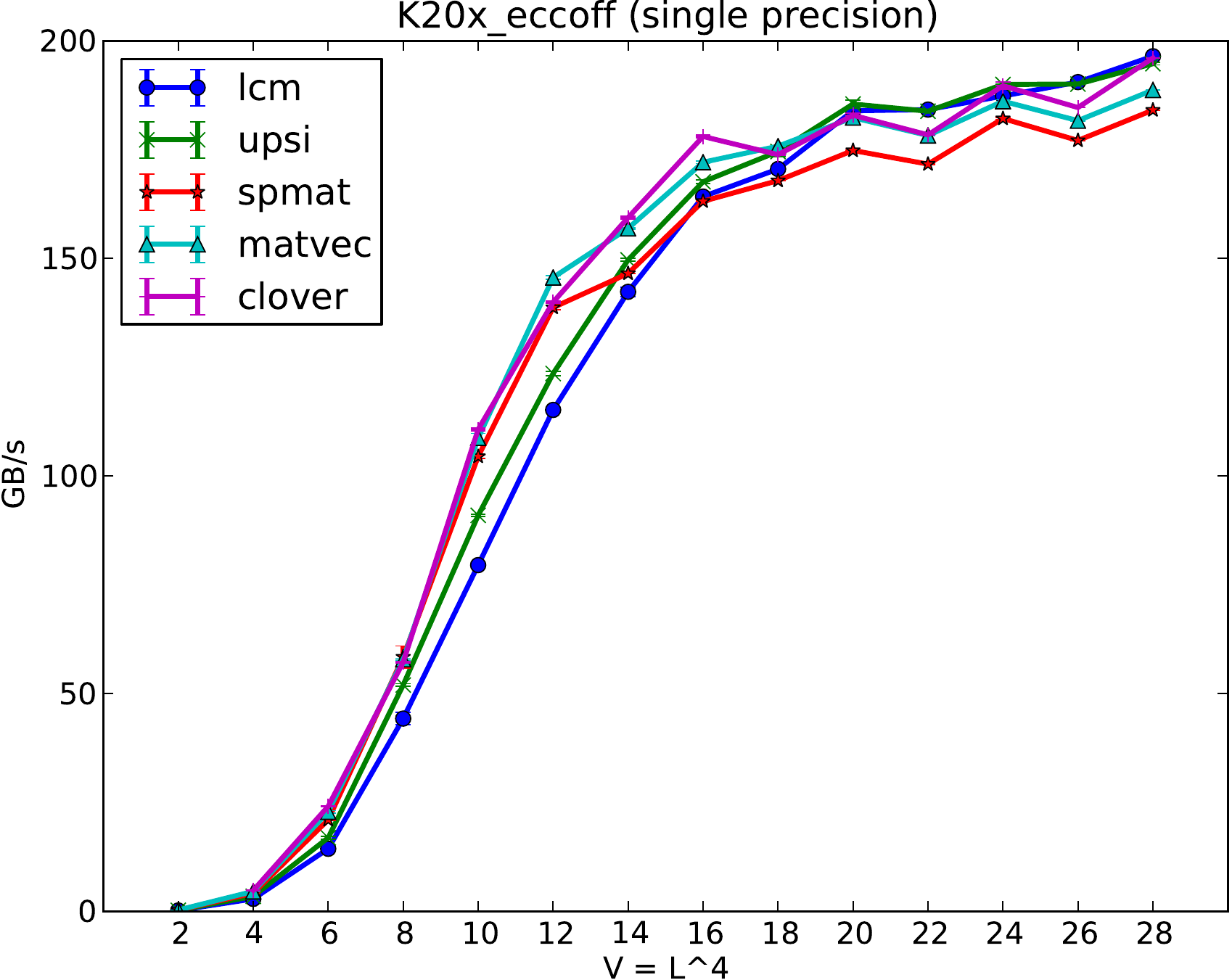}
\caption{\label{fig:gbs_sp}Comparison performance of benchmarking
  kernels in single precision on Tesla K20x (ECC disabled).}
\end{figure}
\begin{figure}[tb]
\centering
\includegraphics[width=1.0\columnwidth]{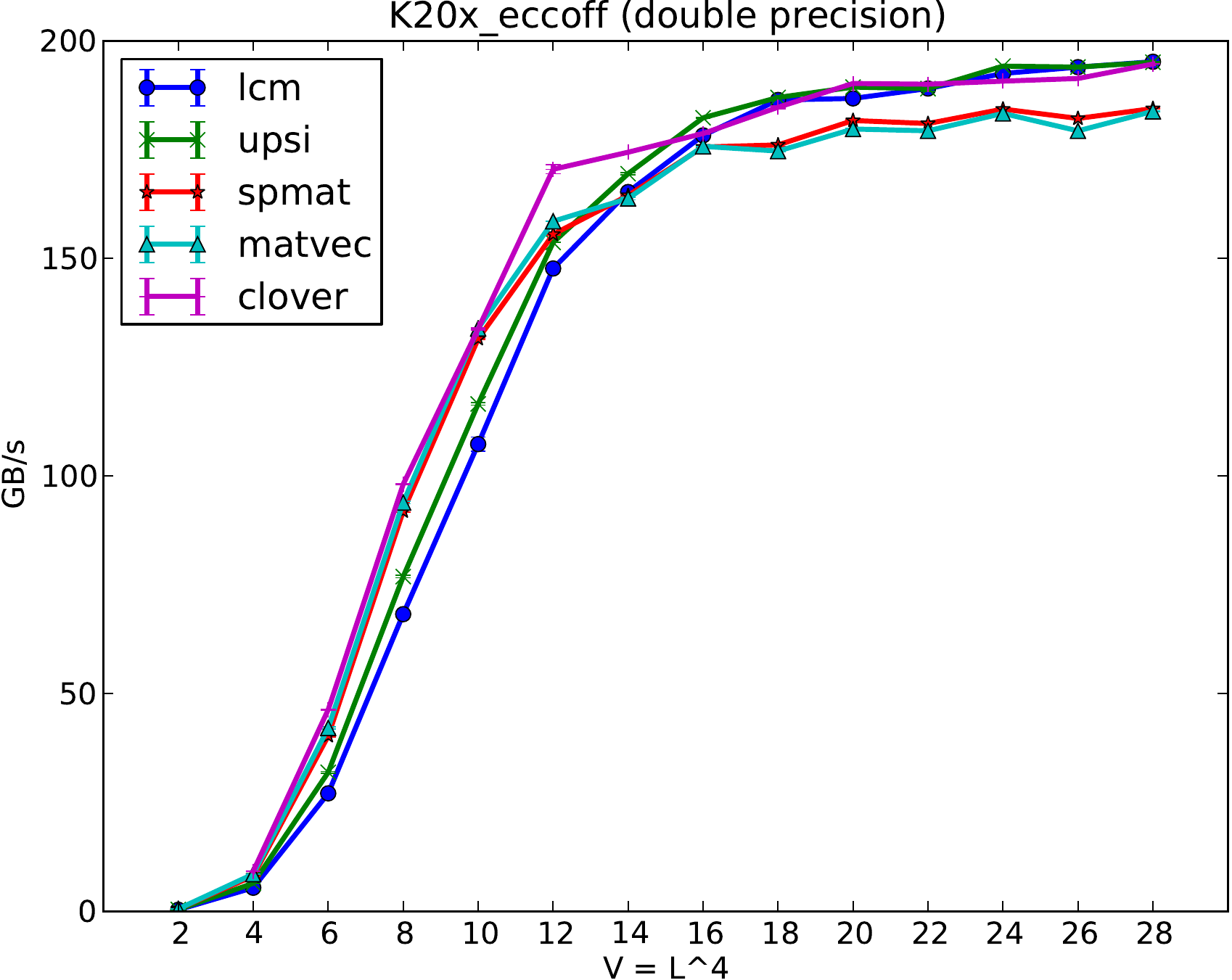}
\caption{\label{fig:gbs_dp}Comparison performance of benchmarking
  kernels in double precision on Tesla K20x (ECC disabled).}
\end{figure}

\begin{figure}[tb]
\centering
\includegraphics[width=1.0\columnwidth]{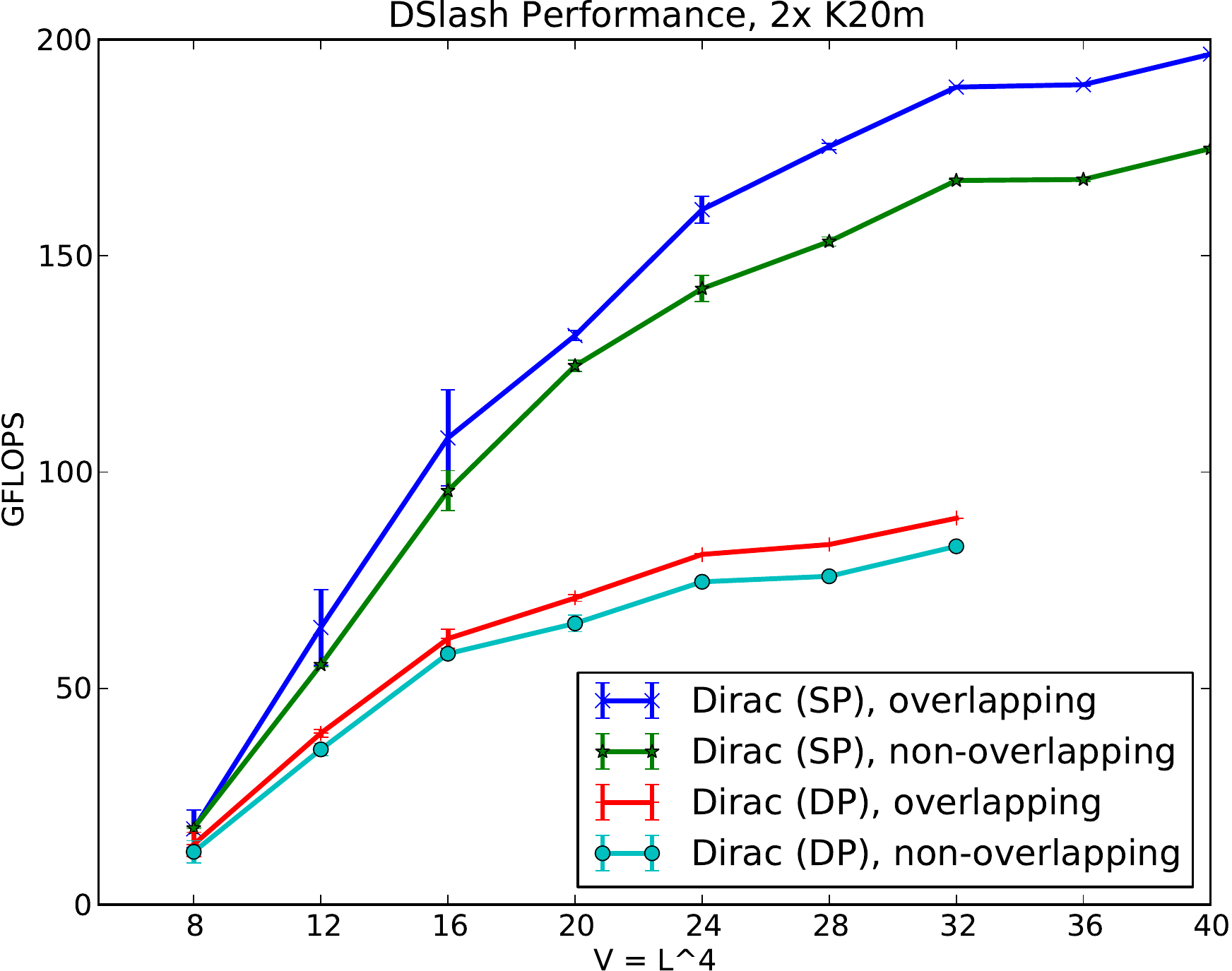}
\caption{\label{fig:overlap_sp}Performance of the hopping part of the
  Wilson Dirac operator. Comparison when overlapping of inter-GPU
  communication and computation is enabled and disabled. Using 1
  NVIDIA K20m per node (ECC enabled).}
\end{figure}

\section{Experimental results}
\label{sec:results}

\subsection{Hardware description}

Our experimental results were acquired on the Jefferson Lab 12k
cluster.
Nodes of this cluster contain dual-socket Xeon E5-2650 CPUs running at
2.0 GHz. The nodes run CentOS 6.2.
We used a node with four NVIDIA K20x GPUs (GK110 architecture)
installed.
The K20x has as a peak performance of 1.3 TFlops (DP) and a peak
memory bandwidth of 250 GB/sec (ECC disabled).
We used the CUDA toolkit version 5.5 and the NVIDIA UNIX x86-64 Kernel
Module version 319.37.
For the 2 GPUs benchmark run detailed in Subsec. \ref{sec:overl} we
used 2 K20m nodes with MVAPICH2 version 1.9 with CUDA-aware MPI.

Our application benchmark on multiple GPUs was run on the XK cabinets
of the NCSA BlueWaters system and on the Titan Supercomputer at the
OLCF.
These nodes comprise 1 AMD 6276 Interlagos Processors (8 physical
cores) and 1 NVIDIA GK110 Kepler accelerator.
The CPU comparison run was deployed on XE nodes which comprise 2 AMD
6276 Interlagos Processors per node.

\subsection{Single GPU Performance}

We note that our test functions (Sec.~\ref{sec:functions}) on the K20x
architecture are memory bandwidth bound in single and double precision
(SP, DP).
Thus, we are interested in the fraction of the memory bandwidth that
our kernels sustain on that hardware vs. the theoretical maximum
memory bandwidth.
However, since the sustained bandwidth and performance are directly
related we will use both synonymously.

Figure~\ref{fig:gbs_sp} and \ref{fig:gbs_dp} show the sustained
bandwidth as a function of the local sub-grid size and for SP and DP
respectively. 
The curves corresponding to different kernels show a universal
behavior and (nearly) fall on top of each other indicating that the
performance of our generated code depends very little on the actual
function which it implements.

In SP and DP for smaller volumes the sustained bandwidth is steadily
increasing with the volume up to a ``shoulder region'' at around
$V=16^4$ for SP and $V=12^4$ for DP from where on the performance
increases only little.
For the largest volumes the kernels sustain about 79\% of the
theoretical maximum memory bandwidth.

The shape of the performance curve is what one would expect.
The GPUs requires many threads resident to a single Streaming
Multiprocessor (SM) in order to effectively hide latency of memory
accesses.
The shoulder region seems to indicate a ``thread saturation'' of SMs.
Below the shoulder region not enough thread blocks are resident and
consequently the sustained memory bandwidth decreases.

\subsection{Overlapping Inter-GPU Communication}

\label{sec:overl}

In the following we demonstrate the impact of overlapping
communication and computation as detailed in
Sec.~\ref{sec:overlapping}.

The hopping part of the Wilson discretization of the Dirac operator
$$
H(x,x')=\sum\limits_{\mu=1}^4
(1-\gamma_\mu)U_x^\mu\delta_{x+\hat\mu,x'} +
(1+\gamma_\mu)U_{x-\hat\mu}^{\mu\dagger}\delta_{x-\hat\mu,x'}
$$
was implemented using the high-level domain abstractions provided by
the QDP-JIT/PTX library.

For this test we used a total of two NVIDIA K20m installed in two
12k compute nodes, i.e., we see overlapping of compute kernels with
MPI communication routed through the PCIe bus and the InfiniBand
interconnect.

Figure~\ref{fig:overlap_sp} shows the combined performance of the
hopping part as a function of the global volume.
Shown is the comparison of the performance when overlapping is enabled
or disabled, in SP and DP.
For SP we measured an 11\% performance increase for the largest
volume.
The impact on DP performance is smaller, about 7\% for the largest
volume.
This indicates that overlapping of MPI communication and CUDA kernel
computation takes place.

We would like to note that this benchmark test was included solely to
demonstrate the effect of overlapping communication and computation.
It is not meant to compete with implementations which focus
optimizations exclusively to this operator.
Our implementation of the hopping term was generated from its
high-level representation.

We measured the same operator as available from the QUDA library
(version 0.6) \cite{quda} on the same hardware. 
We configured QUDA with ``overlapping communications'', targeting
compute capability ``3.5''.
In SP, $V=40^4$ we measured 346 GFLOPS (with uncompressed gauge
fields) where our implementation achieves 197~GFLOPS (speedup factor
$1.76$).
In DP, $V=32^4$ we measured 171 GFLOPS where our implementation
achieves 90~GFLOPS (speedup factor $1.9$).
QUDA can apply other optimizations to the Dslash operator like
reconstruction of SU(3) matrices from 8 or 12 real numbers to gain
additional speedups.
However, we used a set of optimizations such that the same amount of
work was done by QUDA and our implementation since we are focusing
here on communications.
These results give an indication of the amount of ``headroom''
available for hand tuned optimizations, without employing gauge
compression.

\subsection{Hybrid Monte Carlo on Blue Waters}

\begin{figure}[tb]
\centering
\includegraphics[width=1.0\columnwidth]{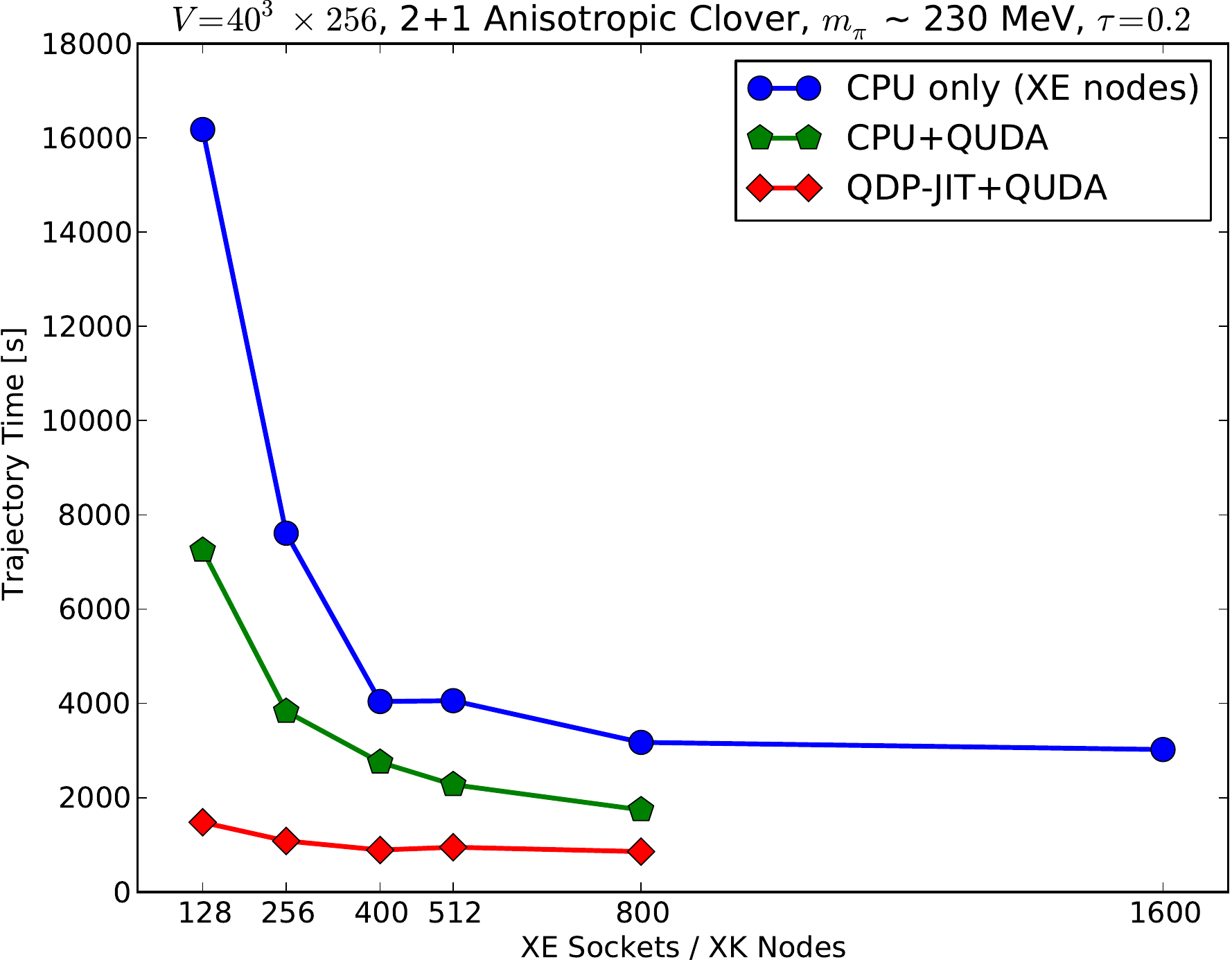}
\caption{\label{fig:HMC}Strong scaling of HMC on Blue Waters.}
\end{figure}

\begin{figure}[tb]
\centering
\includegraphics[width=1.0\columnwidth]{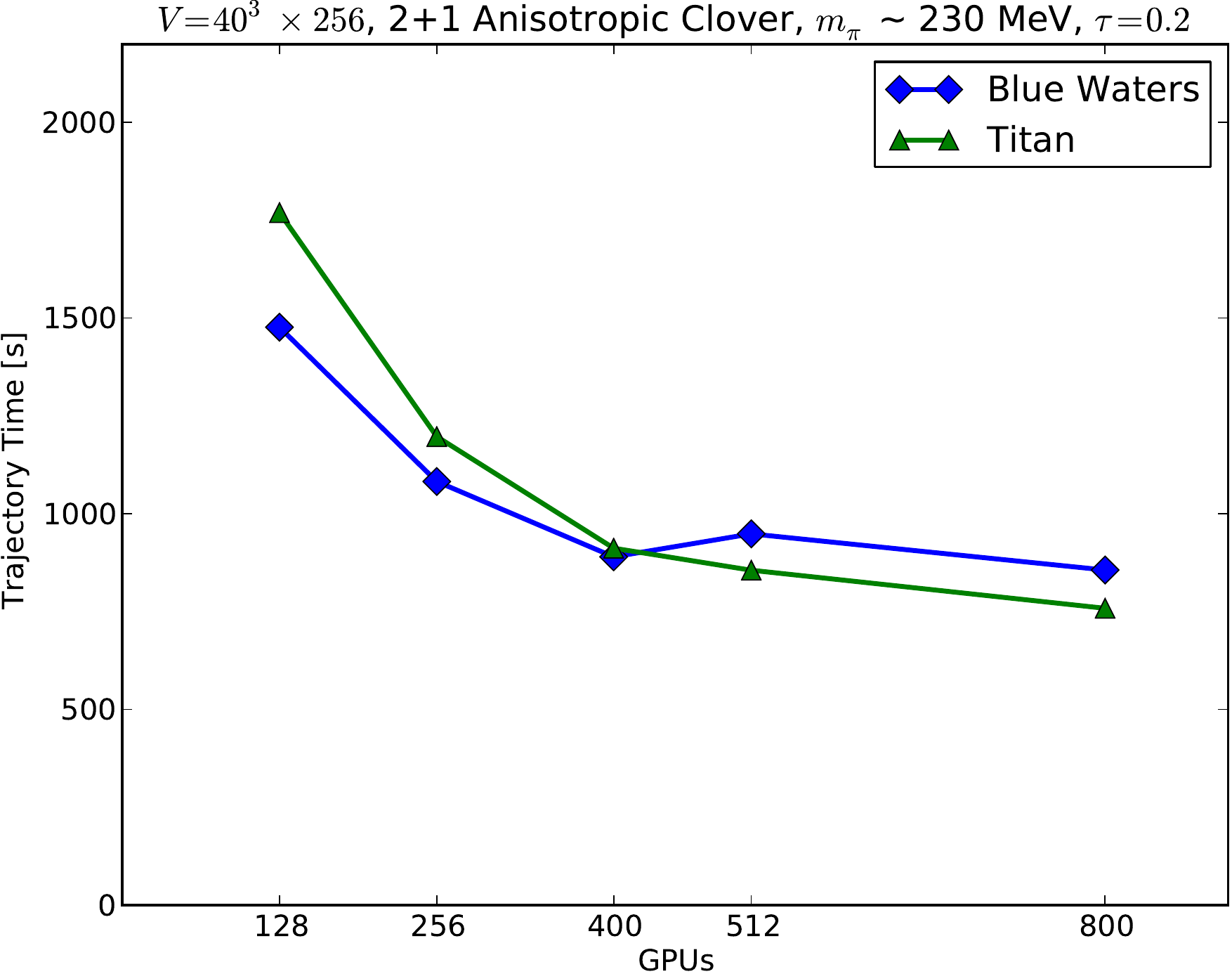}
\caption{\label{fig:Titan}Comparison Blue Waters and Titan. Shown is
  the strong scaling of HMC for the configuration ``QDP-JIT+QUDA''.}
\end{figure}

Our implementation enables us to deploy the full Chroma
gauge-generation program on large scale GPU-based machines.
Using additionally the linear solvers from the QUDA library
\cite{Clark:2009wm} results in a high-performance configuration for
running HMC calculations on systems such as Titan at
the OLCF, and Blue Waters at the NCSA.

We are using QUDA's device interface to call-out from Chroma to the
linear solvers.
The interface supports the optimized data layout as used in
the QDP-JIT/PTX library and thus eliminates the requirement to copy the
spinor, gauge and clover fields to the CPU memory and changing the
data layout prior to calling the solvers.

We will now discuss our scaling results which we obtained using
exactly the same simulation parameters as we are using in our
production program, modulo shortening the trajectory lengths.
The global lattice size is set to $V=40^3\times 256$.
We are running simulations with $N_f=2+1$ dynamical flavors of Wilson
clover \cite{Sheikholeslami1985} fermions on an anisotropic
lattice. We employ mass preconditioning \cite{Hasenbusch:2002ai} and
the rational approximation \cite{Clark:2004cp} to calculate the
determinant of the Dirac operator with the strange quark mass.

Chroma was deployed throughout in double precision on Blue Waters in
several software configurations.
The individual configurations differ by how the GPUs are
utilized for different parts of the calculation.
Figure~\ref{fig:HMC} shows the time for a trajectory for the various
software configurations.

In order to get a baseline of performance we deployed Chroma in a
CPU only configuration.
We varied the number of XE sockets from $128$ to $1600$.
We found good scaling up to 400 XE sockets from where on performance
shows only modest increase; going from 800 to 1600 has only a marginal
effect.

Plugging in the QUDA linear solvers, i.e. configuration ``CPU+QUDA'',
has a positive impact on the performance; the speedup factor over the
whole range of partition sizes stays roughly constant;
we measured a speedup factor of $\sim2.2$ for 128 GPUs/CPU sockets and 
a factor of $\sim1.8$ for 800 GPUs/CPU sockets.
However the overall scaling trend has not improved.
This configuration is sensitive to Amdahl's law effects combined with
the poorer scaling of the GPU part due to repeated copying of data
fields between the CPU and the GPU and changing data layouts.

Turning to a Chroma build over QDP-JIT/PTX resulted in generating
about 200 GPU kernels for this particular trajectory.
Based on our earlier measurements we estimate the total JIT
compilation overhead to be around $\sim10-30$ seconds, and thus negligible
compared to the time needed for the trajectory.

The configuration ``QDP-JIT+QUDA'' achieves the highest performance
of all deployed configurations.
Full benefit is taken from the algorithmic improvements (QUDA GCR
solver) due to the seamless interface between the QUDA and QDP-JIT/PTX
libraries, and the fact that all computation is accelerated effectively
alleviates the effects of Amdahl's law.
A speedup factor of $\sim11.0$ could be achieved on $128$
GPUs/CPU sockets, and a significant factor of $\sim3.7$ on $800$
GPUs/CPU sockets.
When compared to the ``CPU+QUDA'' configuration, ``QDP-JIT+QUDA''
achieves a speedup factor of $\sim2.0$ for 800 GPUs.

Figure~\ref{fig:Titan} shows the timing for the trajectory on Blue
Waters and the Titan supercomputer.
As expected the Blue Waters and Titan results are hardly
distinguishable when bearing in mind that our benchmark timings on
these systems typically show some amount of fluctuation.

The most efficient machine size for the $40^3 \times 256$ lattice size
is 128 XK nodes where the integrated resource costs are 258 vs. 52
node hours for the ``CPU + QUDA'' and ``QDP-JIT + QUDA''
configurations respectively.
Thus, due to this work the computational cost for our HMC simulations
with this lattice size could be reduced by a factor of $\sim5$.

\section{Related Work}
\label{sec:related}

Automatic GPU code generation was addressed, e.g., in
\cite{Chen:2012:AOC:2357488.2357661, ET-CUDA}.
There, however, kernel code is generated using CUDA C++.
This creates a dependency on calling NVCC at runtime.
Previously we used a similar approach 
\cite{Winter:2012wy, Winter:2011dh} and our findings were that
this does not lead to a smooth process.
The NVCC compiler, designed for static compilation, is not fast
in translating code.
Thus caching of compiled kernels on the filesystem is desired.
Also calling NVCC on the compute node, is not possible in all
environments, e.g. on Cray compute node Linux environment.
Previous efforts completely ignore GPU memory management,
optimizations such as optimizing the data layout,
kernel auto-tuning and
overlapping of computation and communication.

Development of an LQCD application using OpenCL was reported in
\cite{Bach20132042}.
All operations involved in an HMC simulation were implemented
separately as kernels.
This work supports single GPUs only and reports sustaining between
77\% and 80\% of the peak memory bandwidth.

Porting (parts of) LQCD calculations to the CUDA architecture was
addressed in previous work 
\cite{Babich:2011:SLQ:2063384.2063478, Babich:2010mu, Clark:2009wm,
  Alexandru:2011:EIO:2060099.2060319, Egri:2006zm, Bonati:2011dv,
  wagner_gtc13}. 
However, these efforts belong to the class of traditional porting as
focus was limited to only specific parts of LQCD calculations.

\section{Conclusions}
\label{sec:conclusions}

The layered structure of the Chroma software architecture allowed us
to provide a reimplementation of the low-level layer and with it to
port the whole application layer to the CUDA architecture.
The reason why this was tractable is that this low-level layer
implements a data-parallel language with the use of expression
templates. 
Those make the structure of operations accessible to compile-time
computations which allowed us to build kernel code generators and
automate the CUDA memory management.
A necessity was the availability of a JIT compiler for the CUDA
architecture.

Due to this work we reduced the computational cost for our production
Hybrid Monte Carlo simulations on Blue Waters and Titan by a factor of
$\sim5$.
While this worked very well for the scientific domain of lattice QCD,
especially for Chroma, we expect the approach can also benefit in
other domains with data-parallel computational patterns.

\section{Future Work}

We are exploring the possibility to interface to a compiler framework
such as LLVM \cite{Lattner:2004:LCF:977395.977673}.
This would allow us to target other architectures as well.

\section{Acknowledgments}

We gratefully acknowledge support for this work on OLCF Titan through
Directors Discretionary Allocation LGT006 (2012-2013) and through the
INCITE project Allocation LGT003 2012-2013.

This research is part of the Blue Waters sustained-petascale computing
project, which is supported by the National Science Foundation (award
number OCI 07-25070) and the state of Illinois. Blue Waters is a joint
effort of the University of Illinois at Urbana-Champaign and its
National Center for Supercomputing Applications.

Partial support for this work was provided through the Scientific
Discovery through Advanced Computing (SciDAC) program funded by
U.S. Department of Energy, Office of Science, Offices of Advanced
Scientific Computing Research, Nuclear Physics and High Energy
Physics.

This research was in part supported by the Research Executive Agency
(REA) of the European Union under Grant Agreement number
PITN-GA-2009-238353 (ITN STRONGnet).

Notice: Authored by Jefferson Science Associates, LLC under U.S. DOE
Contract No. DE-AC05-06OR23177. The U.S. Government retains a
non-exclusive, paid-up, irrevocable, world-wide license to publish or
reproduce this manuscript for U.S. Government purposes.

\section{Code Availability}

The latest version of the QDP-JIT/PTX library is always
available in a publicly-accessible source code repository
\cite{qdp-jit}.


\bibliography{paper}

\end{document}